\documentclass{emulateapj}
\usepackage{epsf}
\usepackage{xspace}
\usepackage{amsmath}
\def\figdir{.}

\def\pgrid{P_{\rm grid}}
\def\delDC{\delta_{\rm DC}}
\def\DelDC{\Delta_{\rm DC}}
\def\auni{a_{\rm uni}}
\def\abox{a_{\rm box}}
\def\dabox{\dot{a}_{\rm box}}

\def\dim#1{\mbox{\,#1}}
\def\figname#1{\figdir/#1}

\def\hide#1{}

\begin{document}

\title{Implementing the DC Mode in Cosmological Simulations with Supercomoving Variables}

\author{Nickolay Y.\ Gnedin\altaffilmark{1,2,3}, Andrey V.\
  Kravtsov\altaffilmark{2,3,4}, and Douglas H.\ Rudd\altaffilmark{5,6}}
\altaffiltext{1}{Particle Astrophysics Center, 
Fermi National Accelerator Laboratory, Batavia, IL 60510, USA; gnedin@fnal.gov}
\altaffiltext{2}{Kavli Institute for Cosmological Physics and Enrico
  Fermi Institute, The University of Chicago, Chicago, IL 60637 USA;
  andrey@oddjob.uchicago.edu} 
\altaffiltext{3}{Department of Astronomy \& Astrophysics, The
  University of Chicago, Chicago, IL 60637 USA} 
\altaffiltext{4}{Enrico Fermi Institute, The University of Chicago,
Chicago, IL 60637}
\altaffiltext{5}{Yale Center for Astronomy \& Astrophysics, Yale University,
New Haven, CT 06520, USA; douglas.rudd@yale.edu}
\altaffiltext{6}{School of Natural Sciences, Institute for Advanced
  Study, Princeton, NJ 08540, USA} 

\begin{abstract}
As emphasized by previous studies, proper treatment of the density
fluctuation on the fundamental scale of a cosmological simulation
volume -- the ``DC mode'' -- is critical for accurate modeling of
spatial correlations on scales $\gtrsim 10\%$ of simulation box
size. We provide further illustration of the effects of the DC mode on
the abundance of halos in small boxes and show that it is straightforward
to incorporate this mode in cosmological codes that use the
``supercomoving'' variables. The equations governing evolution of 
dark matter and baryons recast with these variables are particularly
simple and include the expansion factor, and hence the effect of the DC
mode, explicitly only in the Poisson equation. 
\end{abstract}


\section{Introduction}
\label{sec:intro}

Cosmological simulations have become the main theoretical tool for
studying the evolution of cosmic structures. Their applications range
from modeling the sub-parsec scale environments of first stars and
supermassive black holes to the large-scale distribution of matter
and galaxies. Correspondingly, sizes of simulated regions range from hundreds
of kiloparsecs to gigaparsecs.

Many modern simulations use small simulation volumes, focusing limited
computational resources on resolving important small-scale dynamics of
galaxies or dark matter substructure. In order for such a simulation
to remain a fair realization of a region of the universe, it must
properly account for the time evolution of the non-vanishing
fluctuation of the cosmic density at the scale of the simulation
box. Following \citet{sims:s05}, we call this fluctuation the ``DC
mode,''\footnote{By analogy with the constant electric Direct
  Current.} $\delDC$, to reflect the fact that it is constant in
space over the simulation volume.

While any simulation of a finite volume has a non-vanishing DC
mode, it is not easy to compute for a volume of arbitrary geometry. 
Therefore, in the following we assume that the simulation
volume is a cubic box of size $L$, and that periodic boundary
conditions are imposed along each box dimension. This is indeed the most
common setup for cosmological simulations.

Throughout this paper we use a cosmological model consistent with the third
year WMAP results ($\Omega_m=0.24$, $h_{100}=0.73$, $\sigma_8=0.75$,
and $n_S=0.95$), as this is the model used in our reference
$80h^{-1}\dim{Mpc}$ simulation that we adopt as an approximation to a
representative volume of the universe. None of our conclusions,
however, are dependent on the specific values of cosmological
parameters.

\section{Effects of the DC mode in Cosmological Simulations}
\label{sec:finite}

In the infinite universe the power spectrum of density fluctuations,
$P(k)$, and their correlation function, $\xi(r)$, are the Fourier
transforms of each other. In a cubic periodic simulation box of finite
size this is no longer true. Since the real space quantities are
directly related to observables, one can argue that it is more
important to maintain correct $\xi(r)$ than $P(k)$ within the
simulation volume \citep{sims:p97}. In this case the power spectrum of
the density fluctuations inside the simulation box of size $L$ becomes
a convolution of the true cosmic power spectrum $P(k)$ and the window
function of the simulation volume $W_L(\vec{k})$,
\begin{equation}
  \pgrid(\vec{k}) = \int P(k^\prime)
  W_L(\vec{k}-\vec{k}^\prime) d^Dk^\prime,
  \label{eq:pgrid}
\end{equation}
where
\[
  W_L(\vec{k}) = \prod_{j=1}^D \frac{\sin(k_jL/2)}{\pi k_jL}
\]
and $D$ is the dimension of space. Even if $P(0)=0$, $\pgrid(\vec{0})$
is not, in general, equal to zero because rms
value of density fluctuation at the box scale is not zero. 
Simulations aiming to model density fluctuations on scales comparable to box size correctly must therefore incorporate a non-zero DC mode.

\begin{figure}[t]
\plotone{\figname{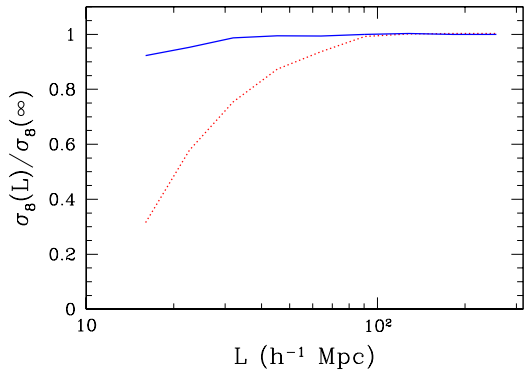}}
\caption{\label{fig:sigma8} The linear mass variance in
  spheres of radius $8h^{-1}\dim{Mpc}$ ($\sigma_8$) at $z=0$  as a function of
  the simulation box size $L$, as measured from a large number of
  realizations. Red dotted line shows the standard method, where the
  true linear $P(k)$ is used to generate the cosmic density field,
  while the blue solid line shows the method that uses $\pgrid$ for
  setting the density amplitude on the grid.}
\end{figure}

In order to illustrate the effect of using $\pgrid$ instead of the
true linear $P(k)$ while generating the cosmological initial
conditions, we create a large number of realizations of the linear
density field for each value of the simulation box size $L$ and
measure the linear mass variance on the scale of $8h^{-1}\dim{Mpc}$,
$\sigma_8$, as the average over the whole ensemble\footnote{The
specific number of realizations, $N$, for each value of $L$ is
determined by the requirement that the average $\sigma_8$ is measured
to 1\% precision.}.  Figure \ref{fig:sigma8} shows $\sigma_8$ as a
function of simulation box size $L$ at $z=0$ for two sets of
realizations - one that uses $\pgrid$ and another one that uses true
linear $P(k)$. In the former method the linear mass variance in
spheres is preserved to at least 90\% at scales as small as half the
box size, while in the standard method
\citep[e.g.,][]{sims:b01,sims:ppap08} the mass variance is only
accurate in spheres that are less than 10\% of the box size in
radius. The same is true for many other real-space clustering
measures, as is amply demonstrated by \citet{sims:p97} and
\citet{sims:s05}.

Using $\pgrid$ to set up initial conditions clearly increases the
fidelity of a small-box cosmological simulation. However, how 
important is it to have a non-zero DC mode? After all, the non-zero value
of $\pgrid(0)$ only implies a non-vanishing \emph{rms} of a
Gaussian-distributed DC mode. A single simulation can always impose
the constraint of $\delDC=0$ without violating the statistical
properties of the initial conditions, although, strictly speaking, such
initial conditions will not be a true random realization of the
universe.  Hence, if an ensemble of simulations is performed,
then proper sampling of the DC mode is crucial, even if the ensemble
only contains just two simulations.

\begin{figure}[t]
\plotone{\figname{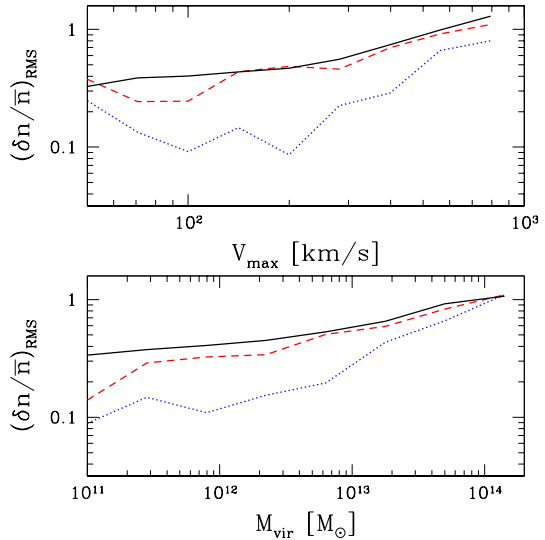}}
\caption{\label{fig:mf} The rms fluctuations in the numbers of dark
  matter halos as a function of halo maximum circular velocity (top)
  or the virial mass (bottom) in cells of
  $L=20h^{-1}\dim{Mpc}$. Dotted blue and dashed red lines show the
  average over two ensembles of 5 simulations each, with the DC mode
  forced to zero and the properly computed DC mode respectively.
  Simulations followed evolution in a box of $L=20h^{-1}\dim{Mpc}$;
  each of the ten simulations used different random realizations of
  initial conditions. Solid black lines show the rms computed from a
  single simulation of the same cosmology of $L=80h^{-1}\dim{Mpc}$ box
  size.}
\end{figure}

To illustrate this point, we show in Figure \ref{fig:mf} the rms
variation in the number of dark matter halos as a function of their
maximum circular velocity or the virial mass. To compute the rms, we
generate two sets of five different random realizations of initial
conditions, in a $L=20h^{-1}\dim{Mpc}$ box. In the first set of
initial conditions the DC mode is forced to be zero, while in the
second set the DC mode is computed properly as a Gaussian distributed
random number with the rms value of $\pgrid(0)/L^3$. Each of the ten
realizations of initial conditions is then evolved to $z=0$ with the
N-body part of the Adaptive Refinement Tree (ART) code
\citep{misc:kkk97,misc:k99} with $128^3$ particles and $2^{12}$
dynamic range.

As a control sample, we use a halo population from a single simulation
of $L=80h^{-1}\dim{Mpc}$ with zero DC mode. The simulation followed
$512^3$ particles with a dynamic spatial range of $2^{15}$. Simulation
results of this larger box at $z=0$ are used to compute the rms number
density fluctuations of dark matter halos in cubic boxes of
$20h^{-1}\dim{Mpc}$ on a side. In both simulations halos were
identified using a variant of the Bound Density Maxima algorithm, as
described in \citet{kravtsov_etal04}.

As Figure~\ref{fig:mf} illustrates, the ensemble of simulations with
the DC mode (red dashed lines) properly accounts for the fluctuations
on the box size scale. Setting the DC mode to zero (blue dotted lines)
results instead in a factor of $\sim 2-3$ underestimate of the true
variance.

\section{Incorporating the DC Mode in a Cosmological Code}
\label{sec:dcmode}

\begin{figure}[t]
\plotone{\figname{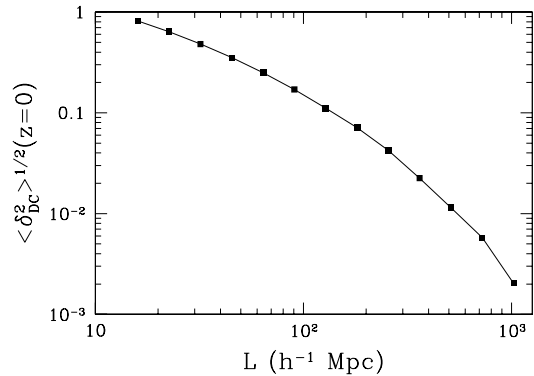}}
\caption{\label{fig:dcmode} The rms amplitude of the DC mode at $z=0$
  as a function of the simulation box size $L$.}
\end{figure}
The DC mode is substantial for many of the commonly used simulation
box sizes.  For example, Figure \ref{fig:dcmode} shows the rms DC mode
as a function of the simulation box size. The rms DC mode falls below
1\% only for box sizes $L>500h^{-1}\dim{Mpc}$. While $1\%$ may seem
like a small number, many studies use such boxes to calibrate
statistics, such as power spectrum and abundance and clustering of
halos, to accuracy of $<5\%$. Therefore, effect of the DC mode may
need to be evaluated even for boxes of hundreds of megaparsecs in
size.

The DC mode thus cannot be ``approximately neglected,'' at least not
in any study aiming at obtaining correct statistics on the scale of
the simulation box size.  How easy it is to incorporate the DC mode
depends on the nature of a cosmological simulation code and could be a
non-trivial task.

Cosmological simulations most commonly employ periodic boundary
conditions and simulation box can be considered as a separate universe
that expands at a different rate than the target model
universe. Following \citet{sims:s05}, we define two expansion factors:
the scale factor of the simulation box, $\abox$, and the true, global
expansion factor of the universeas, $\auni$. As mass is conserved, the
average mass density in the simulation box should be equal to the
density of a region with the overdensity $\delDC$ in the target model
universe,
\begin{equation}
  \frac{\Omega_M}{\abox^3} = \frac{\Omega_M}{
  \auni^3}\left[1+\delDC(\auni)\right],
  \label{eq:den}
\end{equation}
or
\begin{equation}
  \abox = \frac{\auni}{\left[1+\DelDC D_+(\auni)\right]^{1/3}},
  \label{eq:abox}
\end{equation}
where we explicitly spelled out the dependence of the DC mode on the
universal expansion factor $a_{\rm uni}$, $\delDC(\auni) \equiv \DelDC
D_+(\auni)$. Here $D_+$ is the linear growth factor of density
perturbations; it is one (namely, growing) of the two independent
solutions of the linear perturbation equation for dust-like matter in
the Newtonian approximation\footnote{
One may question the validity of the Newtonian approximation on large
scales in simulations of large box sizes, $L=O(c/H_0)$, given that
evolution of large-scale modes should be governed by fully
relativistic equations. However, we note that equation (\ref{eq:growth})
is identical to the corresponding fully relativistic linear
perturbation growth equation in the synchronous gauge or, more
generally, in the comoving ``total matter gauge''. The
equations governing evolution of perturbations in the synchronous
gauge are presented, for example, in \citet{ma_bertschinger95}. We can
derive equation identical to eq.~(\ref{eq:growth}) above by 1)
substituting their eq. 21a into 21c, 2) considering the case of
universe with energy density dominated by non-relativistic matter
($\delta T_0^0=\bar{\rho}\delta$, $\delta T_i^i=0$), for which
$dh/d\tau=-2\delta$ (their eq. 42), and 3) by making appropriate
transformations from conformal time $\tau$ to physical time $t$ using
$d\tau=dt/a(\tau)$.  This means that linear evolution of
perturbations is treated correctly in the Newtonian approximation of
cosmological codes at any scale, provided one interprets results of
simulations in the appropriate gauge \citep[see,
e.g.,][]{wands_slosar09,chisari_zaldarriaga11}.} \citep[e.g.,][]{bonnor57,peebles80}:
\begin{equation}
\ddot{\delta} + 2\frac{\dot{a}}{a}\dot{\delta} =
                           4\pi G \bar{\rho} \delta,
\label{eq:growth}
\end{equation}
where $\delta\equiv(\rho-\bar{\rho})/\bar{\rho}$ is the matter overdensity
with respect to the mean density of the universe $\bar{\rho}$, $a(t)$
is the expansion factor, and derivatives are taken with respect to the
physical time $t$.

We adopt a normalization for $D_+$ such that in a universe filled with
matter and radiation only (i.e., at sufficiently early times),
\begin{eqnarray}
  D_+(a) & = & a + \frac{2}{3}a_{\rm eq} + \frac{a_{\rm
  eq}}{2\ln(2)-3}}\times{\nonumber \\ & &
  \left[2\sqrt{1+x}+\left(\frac{\displaystyle 2}{\displaystyle
  3}+x\right)\ln\frac{\sqrt{1+x}-1}{\sqrt{1+x}+1}\right],
\end{eqnarray}
where $x \equiv a/a_{\rm eq}$ and $a_{\rm eq} \equiv \Omega_R/\Omega_M$
is the scale factor of matter-radiaton equality. The last two terms
can often be neglected, the third one falls below 1\% for $z \la
1{,}100$ and the second one is below 1\% for $z \la 35$.

\citet{sims:s05} also presents a different form for the $\abox -
\auni$ relation, which he calls a ``Lagrangian viewpoint'' as opposite
to the ``Elurian viewpoint'' of equation (\ref{eq:abox}), but in
essence is just a first order expansion of Equation (\ref{eq:abox}) in
$\DelDC$,
\begin{equation}
  \abox\approx \auni\left[1-\frac{1}{3}\DelDC D_+(\auni)\right].
  \label{eq:aboxlag}
\end{equation}
When we use this to compute the rms variation in the number of dark
matter halos shown in Figure \ref{fig:mf}, we find a result which is
virtually indistinguishable from the ``Eulerian viewpoint'' shown with
red dashed lines. We choose to use the ``Eulerian viewpoint'' as our
primary method for accounting for the DC mode, because it explicitly
conserves mass to all orders in the perturbation theory.

The parameter $\DelDC$ is constant in a given simulation and describes
the amplitude of the DC mode in a given realization of initial
conditions. In principle, equation (\ref{eq:abox}) is also valid in
the non-linear regime, if we treat $D_+$ as a specific non-linear
growth mode in a given simulation box that properly accounts for the
coupling of the DC mode to other modes (including modes with
wavelengths larger than $L$). In practice, however, there exists no
way to compute the non-linear growth rate besides the numerical
simulation itself, so hereafter we assume that the DC mode remains
with sufficient precision in the linear regime throughout the whole
time-span of the numerical simulation. Of course, the term
``sufficient precision'' depends on the required precision of the
simulation results. For example, for our ensembles of
$20h^{-1}\dim{Mpc}$ boxes the rms value of $\delDC(1)$ is almost 0.7;
never-the-less, the linear approximation for the DC mode is sufficient
to achieve about 20\% agreement between an ensemble of
$20h^{-1}\dim{Mpc}$ simulations and a single $80h^{-1}\dim{Mpc}$ run
in Figure \ref{fig:mf}.

Historically, the DC mode was sometimes incorporated into a simulation
by appropriately rescaling the values of cosmological parameters. An
example of such rescaling is also given in the appendix of
\citet{sims:s05}. Effectively, in a spatially flat cosmology such a
rescaling introduces non-zero spatial curvature (or additional
curvature if the universe is assumed to be spatially non-flat) with
the extra curvature parameter $\Delta \Omega_K \approx -(5/3)\Omega_M
\DelDC$.

However, such approach does not work in more general cosmological
models. For example, one \emph{cannot} consider simulation with a DC mode as
a separate universe in models with non-negligible amount of
relativistic matter, general dark energy component, decaying dark
matter, or modified gravity.

We, therefore, advocate an alternative approach that can be used to
include the DC mode in simulation codes exactly. This can be done by
following both $\abox$ and $\auni$ as a function of cosmic time and
incorporating this difference in simulation equations explicitly. A
disadvantage of this approach is that some of the equations become
unnecessarily complicated. For example, the evolution equation for the
peculiar velocity $\vec{v}_{\rm pec}$ of a dark matter particle in the
expanding coordinate system commonly used in cosmological codes is
\[
  \frac{d\vec{v}_{\rm pec}}{dt} = - \frac{\dabox}{\abox} \vec{v}_{\rm pec} -
  \frac{1}{\abox} \nabla_x\phi,
\]
where dot symbolizes the time derivative, $\phi$ is the peculiar
gravitational potential, and spatial derivatives are taken with
respect to the comoving coordinates $\vec{x}$, which is emphasized by
the subscript $x$ under the gradient symbol. Computing $\dabox$
requires a differentiation of equation (\ref{eq:abox}) and leads to
uncouth and inelegant (but fully tractable) formulae.

A much more elegant way to incorporate the DC mode in a cosmological
code is via the so-called \emph{supercomoving variables}\footnote{The
term ``supercomoving'' was coined by \citet{sims:ms98}.}, first
introduced by \citet{sims:dkps80}, and then used in several numerical
works
\citep{sims:s80,sims:ssm83,sims:ss85,ng:g95,sims:ykkk97,sims:ms98,misc:kkh02,sims:t02}.
Specifically, the proper coordinates $\vec{r}$, velocity $\vec{u}$,
density $\rho$, and gravitational potential $\Phi$ are replaced with
comoving coordinates $\vec{x}$, supercomoving peculiar velocities
$\vec{v}$ (to be distinguished from the normal peculiar velocities
$\vec{v}_{\rm pec}$), comoving density $\varrho$, and peculiar supercomoving
gravitational potential $\varphi\equiv a^2\phi$ according to the
following transformations:
\begin{eqnarray}
  \vec{r} & = & a\vec{x}, \label{eq:scbeg} \\
  \vec{u} & = & aH\vec{x} + \frac{1}{a}\vec{v}, \\
  \rho    & = & \frac{\varrho}{a^3}, \\
  \Phi    & = & -\frac{a\ddot{a}}{2}x^2 + \frac{1}{a^2}\varphi.
\end{eqnarray}
In addition, the cosmic time $t$ is replaced with the supercomoving
time, $\tau$, defined such that
\begin{equation}
  d\tau = \frac{dt}{a^2}.
  \label{eq:scend}
\end{equation}
In variables $\tau,\vec{x},\vec{v},\varrho,\varphi$ the equations of
motion of dark matter particles assume the form identical to that in
the proper, non-expanding reference frame, without any explicit cosmological
terms,
\begin{eqnarray}
  \frac{d\vec{x}}{d\tau} & = & \vec{v}, \nonumber \\
  \frac{d\vec{v}}{d\tau} & = & -\nabla_x\varphi. \nonumber
\end{eqnarray}
Similarly, the Euler equations for monatomic gas (polytropic index
$\gamma=5/3$) include no explicit cosmological terms either. The only
equation, in which the cosmic scale factor $a$ appears explicitly, is the
Poisson equation,
\begin{equation}
  \nabla_x^2 \varphi = 4\pi G a (\varrho-\bar{\varrho}) \equiv 4\pi G a
  \bar{\varrho}\delta, 
  \label{eq:poi}
\end{equation}
where $\bar{\varrho}$ is the mean comoving matter density of the
universe. Note that $\bar{\varrho}=\dim{const}$ in cosmological models
in which the dark matter does not decay into radiation.

Incorporating the DC mode into a simulation code using the
supercomoving variables is straightforward: all that is required is to
identify the scale factor $a$ in equation (\ref{eq:poi}) with the box
scale factor $\abox$,
\begin{equation}
  \nabla_x^2 \varphi = 4\pi G \abox \bar{\varrho}\delta,
  \label{eq:poidc}
\end{equation}
and compute both $\abox$ and $\auni$ as a function of the supercomoving 
time $\tau$. 

For simulations including gas and other physics, $\abox$ should be
used anywhere a code would use the scale factor internally. For
example, in unit conversions, in cooling functions and star formation
and feedback recipes, for checking energy conservation, in the
radiative transfer solver, etc.  For some physical processes, for
example redshift-dependent cosmic backgrounds (radiation, cosmic rays,
etc), the proper choice of scaling is model dependent.  If a
hydrodynamic simulation uses cooling rates that account for the cosmic
ultraviolet background in a tabulated form
\citep[e.g.,][]{sims:k03,sims:wss09}, the table should typically be
evaluated at $\auni$.  If the sources of the radiation are
inhomogeneous on the scale of the simulation, however, evaluating at
$\abox$ would more accurately capture the large-scale density
dependence.  The only remaining use for $\auni$ is in identifying the
snapshot epoch and the final epoch (if a simulation is evolved to
$z=0$, the stopping criterion is $\auni=1$, not $\abox=1$).

\section{Conclusions}
\label{sec:conclusions}

In this short note we illustrated a known but not widely appreciated
fact that taking the DC mode into account in small box cosmological
simulations is a requirement for the simulation to serve as a
representative volume of the universe and to model density
fluctuations correctly. The DC mode must be included in simulations if
more than a single realization of the initial conditions is used.

While approximate methods for including the DC mode 
have been proposed, here we advocate the use of the supercomoving
variables in cosmological simulations with which the effect of the DC
mode is limited to a simple multiplicative term in the Poisson
equation (\ref{eq:poidc}) and can be incorporated simply and exactly.

An efficient and reliable cosmology module that computes several
important quantities ($\abox$, $\auni$, supercomoving time $\tau$, the
linear growth rate $D_+$, etc.) via lookup tables in the linear regime
is available from the authors upon request.

\acknowledgements 

We would like to thank Wayne Hu, Robert Wald, and Stephen Green for
useful discussions and the anonymous referee for constructive
criticism. The authors are indebted to Marcel Zemp for finding
numerous typos in the original manuscript. This work was supported in
part by the DOE at Fermilab, by NSF grants AST-0507596 and
AST-0807444, and by the Kavli Institute for Cosmological Physics at
the University of Chicago. The simulations used in this work have been
performed on the Joint Fermilab - KICP Supercomputing Cluster,
supported by grants from Fermilab, Kavli Institute for Cosmological
Physics, and the University of Chicago. This work made extensive use
of the NASA Astrophysics Data System and {\tt arXiv.org} preprint
server.

\eject

\bibliographystyle{apj}
\bibliography{ng-bibs/sims,ng-bibs/cosmo,ng-bibs/misc,ng-bibs/self,ak}

\end{document}